\documentclass[twocolumn,showpacs,preprintnumbers,amsmath,amssymb,superscriptaddress,prl]{revtex4}

\usepackage{graphicx}
\usepackage{dcolumn}
\usepackage{bm}

\begin{document}


\title{Local Tunneling Spectroscopy across a Metamagnetic Critical Point \\in the Bi-layer Ruthenate Sr$_3$Ru$_2$O$_7$}

\author{K. Iwaya}
 \altaffiliation{Present address: Dept. of Physics and Astronomy, University College London, WC1E 6BT London, UK.}
\affiliation{%
RIKEN (The Institute of Physical and Chemical Research), Wako, Saitama 351-0198, Japan\\
}%
\affiliation{%
CREST, Japan Science and Technology Agency (JST), Kawaguchi, Saitama 332-0012, Japan\\
}%
\author{S. Satow}%
\affiliation{%
RIKEN (The Institute of Physical and Chemical Research), Wako, Saitama 351-0198, Japan\\
}%
\affiliation{%
Department of Advanced Materials, University of Tokyo, Kashiwa, Chiba 277-8651, Japan\\
}%
\author{T. Hanaguri}
\affiliation{%
RIKEN (The Institute of Physical and Chemical Research), Wako, Saitama 351-0198, Japan\\
}%
\affiliation{%
CREST, Japan Science and Technology Agency (JST), Kawaguchi, Saitama 332-0012, Japan\\
}%
\author{N. Shannon}
\altaffiliation{Permanent address: H H Wills Physics Laboratory, University of Bristol, BS8 1TL Bristol, United Kingdom.}
\affiliation{%
RIKEN (The Institute of Physical and Chemical Research), Wako, Saitama 351-0198, Japan\\
}%
\affiliation{%
CREST, Japan Science and Technology Agency (JST), Kawaguchi, Saitama 332-0012, Japan\\
}%
\affiliation{%
Department of Advanced Materials, University of Tokyo, Kashiwa, Chiba 277-8651, Japan\\
}
\author{Y. Yoshida}
\author{S. I. Ikeda}
\affiliation{%
Nanoelectronics Research Institute, AIST, Tsukuba, 305-8568, Japan\\
}
\author{J. P. He}
\author{Y. Kaneko}
\affiliation{%
Spin Superstructure Project(ERATO-SSS), JST, AIST Central 4, Tsukuba, 305-8562, Japan\\
}
\author{Y. Tokura}
\affiliation{%
Spin Superstructure Project(ERATO-SSS), JST, AIST Central 4, Tsukuba, 305-8562, Japan\\
}
\affiliation{%
Department of Applied Physics, University of Tokyo, Tokyo, 113-8656, Japan\\
}
\author{T. Yamada}
\affiliation{%
RIKEN (The Institute of Physical and Chemical Research), Wako, Saitama 351-0198, Japan\\
}%
\affiliation{%
CREST, Japan Science and Technology Agency (JST), Kawaguchi, Saitama 332-0012, Japan\\
}%
\author{H. Takagi}
\affiliation{%
RIKEN (The Institute of Physical and Chemical Research), Wako, Saitama 351-0198, Japan\\
}%
\affiliation{%
CREST, Japan Science and Technology Agency (JST), Kawaguchi, Saitama 332-0012, Japan\\
}%
\affiliation{%
Department of Advanced Materials, University of Tokyo, Kashiwa, Chiba 277-8651, Japan\\
}

\date{\today}

\begin{abstract}
The local spectroscopic signatures of metamagnetic criticality in Sr$_3$Ru$_2$O$_7$ were explored using scanning tunneling microscopy (STM).
Singular features in the tunneling spectrum were found close to the Fermi level, 
as would be expected in a Stoner picture of itinerant electron metamagnetism. 
These features showed a pronounced magnetic field dependence across the
metamagnetic critical point, which cannot be understood in terms of a naive
Stoner theory. In addition, a pseudo-gap structure was observed over several tens of meV, accompanied by a $c(2\times2)$ superstructure in STM images. 
This result represents a new electronic ordering at the surface in the absence of any measurable surface reconstruction.
\end{abstract}


\maketitle

The subtle interplay between spin, charge and orbital degrees of freedom in strongly correlated electron systems gives rise to a wide variety of competing electronic phases~\cite{Tokura}. A variety of exotic properties have been reported as a result of this competition between different forms of order, and the associated (quantum) critical behavior. Recently, the bi-layer ruthenate Sr$_3$Ru$_2$O$_7$, $n = 2$ member of the Ruddlesden-Popper series Sr$_{n+1}$Ru$_n$O$_{3n+1}$, has attracted considerable interest because of its criticalities. In dimensionality, Sr$_3$Ru$_2$O$_7$ is intermediate between the $n = 1$ compound Sr$_2$RuO$_4$, a spin triplet superconductor~\cite{Mackenzie}, and the $n = \infty$ system SrRuO$_3$, an itinerant ferromagnet~\cite{Allen}. In common with other Sr$_{n+1}$Ru$_n$O$_{3n+1}$ compounds, four $4d$-electrons are accommodated in its almost triply degenerate $t_{2g}$ orbitals, implying that orbital degrees of freedom are likely to play an important role.

The ground state of Sr$_3$Ru$_2$O$_7$ is a (paramagnetic) Fermi liquid. However, the effective mass of the conduction electrons is greatly enhanced, and there is strong evidence of underlying criticality towards electronically and magnetically ordered states. The magnetic susceptibility shows a well-defined peak at $T \sim 16$~K, and its large value in the limit $T \rightarrow 0$ suggests a substantial Stoner enhancement~\cite{Ikeda}. These results indicate that the system is very close to a ferromagnetic instability. Indeed, a weakly first-order metamagnetic phase transition occurs in an applied magnetic field of 5.5 T for $B // ab$, and 8 T for $B // c$~\cite{Perry1, Grigera1}. Furthermore, at only 5\% doping of Mn impurities on Ru sites, the system becomes an antiferromagnetic insulator and is believed to exhibit orbital ordering~\cite{Mathieu}.

The criticalities should manifest themselves as anomalies in low energy spin and charge excitations. The existence of low-lying ferromagnetic spin fluctuations near the metamagnetic critical point has been confirmed by NMR~\cite{Kitagawa}. Similarly, quasi-particle mass anomalies have been observed in de Haas van Alphen (dHvA) measurements at the critical point~\cite{Borzi}. However, until now, no spectroscopic data spanning the transition have been available. Given that criticality is well defined only for temperatures of a few K, meV (or better) resolution is necessary to capture the spectroscopic signatures of criticality. Scanning tunneling microscopy/spectroscopy (STM/STS) is an ideal probe for this purpose.

The single crystals were grown using a floating zone technique and had a residual resistivity 
of $\sim 7$~$\mu\Omega$cm. A metamagnetic anomaly in Ref.~\cite{Perry1} was clearly observed at $\sim 8$~T for $B // c$ in magnetization measurements, indicating that the crystals are of sufficiently high purity.  The samples were cleaved along (001) planes at $T \sim 80$~K in ultrahigh-vacuum (UHV) conditions and transferred to the STM immediately after cleaving. The weakest bond in the crystal is that between adjacent double layers. We therefore believe that the exposed surface layer after cleaving is the neutral SrO layer. STM/STS measurements were performed at $T = 560$~mK. Differential conductance $dI/dV$ spectra, which measure the local density of states (LDOS), were obtained using a standard lock-in technique. Magnetic fields were applied along $c$-axis to tune the metamagnetic criticality.

Fig. 1(a) shows a typical constant-current STM image of Sr$_3$Ru$_2$O$_7$ at a sample-bias-voltage of $+ 100$~mV. A square atomic lattice is clearly observed with an atomic spacing of $\sim 4$~\AA, which is consistent with the tetragonal lattice constant of the system. The surface crystal structure in the $ab$ plane is shown for comparison in Fig. 1(b)~\cite{Huang}. Comparing these two structures, as far as symmetry and atomic spacing are concerned, the bright spots in the STM image can be explained equally well by Ru/apical oxygen or Sr atomic sites. However, we believe that the electronic states responsible for the tunneling originate from the orbital states of Ru and (apical) oxygen, since the weakly hybridized Sr$^{2+}$ ions contribute a negligible amount of spectral weight to the density of states (DOS) at this bias voltage~\cite{Singh}. 

\begin{figure}
\includegraphics[width=7.5cm]{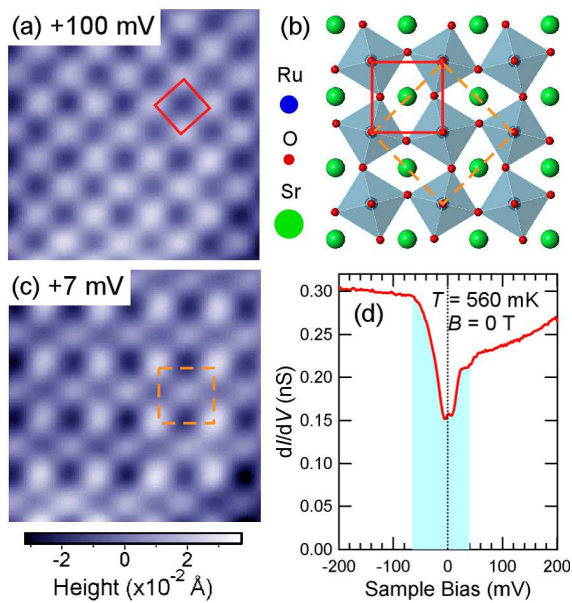}
\caption{(color online)
(a) STM image of Sr$_3$Ru$_2$O$_7$ at $T = 560$~mK (24~\AA $\times$ 24~\AA, 
a sample-bias-voltage $V_s = +100$~mV and a setpoint tunneling current $I_t \sim 50$~pA). The atomic spacing is consistent with the Ru-Ru distance shown as a solid square.
(b) Crystal structure of Sr$_3$Ru$_2$O$_7$ in $ab$ plane. 
Both 1st (SrO plane) and 2nd layer (RuO$_2$ plane) after cleavage are shown. The $c(2\times2)$ unit cell is shown as a dashed square. 
(c)	STM image at $V_s = +7$~mV (24~\AA $\times$ 24~\AA, $I_t \sim 50$~pA). This image was taken at the same location on the sample as (a). The unit cell is shown in the same way as (b). Both images are low-pass filtered to remove high frequency noise.
(d) $dI/dV$ spectrum measured at $T = 560$~mK, $B = 0$~T ($V_{\rm mod}=500~\mu {\rm V}_{\rm rms}$, $f_{\rm mod} = 717$~Hz).
}
\label{fig_1}
\end{figure}

The images at high bias-voltages draw on electrons over a wide energy range and therefore accurately represent the lattice. In this sense, the observation of a uniform square lattice indicates the absence of a surface reconstruction of atomic position. At lower bias-voltage, however, the image becomes more sensitive to the electronic states near the Fermi level $E_F$. A $c(2\times2)$ "rock-salt" type modulation is observed in topographic images for bias-voltages $-60\,{\rm mV} < V_s < +40\,{\rm mV}$ (Fig. 1(c)), accompanied by a pseudo-gap structure in the $dI/dV$ spectrum (Fig. 1(d)). Since all of these features are observed at several different tunneling junction resistances, we can exclude the possibility of a tip-induced effect. 
We note that a similar pseudo-gap~\cite{Barker} and superstructure~\cite{Barker,Matzdorf} have been reported in Sr$_2$RuO$_4$. 
In Sr$_2$RuO$_4$, however, such a superstructure was observed even at high bias-voltages (up to -750 mV), indicating a complete reconstruction of
atomic position~\cite{Matzdorf}. 

This low-energy superstructure can be attributed to an electronic order induced at the surface - for example, a spin or charge density wave, with or without orbital ordering. The idea of a purely electronic reconstruction at a surface has been discussed previously~\cite{Hesper} and Sr$_3$Ru$_2$O$_7$ might provide one of the first examples of such a state~\cite{CDW}. It is clear from bulk probes that the bi-layer ruthenate exists near to spin/charge/orbital ordered states. Such a criticality may be further enhanced at the surface. With respect to the crystal structure, a $c(2\times2)$ superstructure {\it is} also realized in bulk associated with the alternating in-plane rotation of RuO$_6$ octahedra about the (001) axis. However, the two different rotated RuO$_6$ octahedra are electronically equivalent, and the Ru, apical oxygen, and Sr ions cannot be distinguished by symmetry arguments alone. Therefore, the superstructure at low bias-voltages cannot be explained in this way. At this stage, we cannot rule out the possibility that the electronic ordering observed at the surface also occurs in the bulk, but with a modulation too small to be detected by, e.g. x-ray diffraction studies. 

Lowering the bias-voltage still further, we begin to see the signatures of metamagnetic criticality in the LDOS. Metamagnetic transitions in itinerant systems are usually modeled using a simple Ginzburg-Landau ansatz for the free energy, which can in turn be parameterized from a Stoner-type microscopic theory~\cite{Yamada}. In this picture, the transition arises as a result of a sharp peak or upturn in the DOS near $E_F$. The metamagnetism is most pronounced where the Fermi energy lies at a (local) minimum of DOS in zero field. In the most naive, single-particle terms, the metamagnetic transition occurs because the Zeeman splitting of these peaks transfers extra spectral weight to $E_F$. Once the DOS at $E_F$ exceeds the limit set by the Stoner criterion, the system undergoes a metamagnetic transition into a state with an enhanced magnetization. Since in Sr$_3$Ru$_2$O$_7$ the critical field is about 8~T ($B // c$), any spectroscopic signature of Stoner-type metamagnetism should show up on an energy scale of $\sim 1$~meV, for temperatures lower than 10~K.

To probe the fine structure on this energy scale, we measured $dI/dV$ spectra with an energy resolution of 80~$\mu {\rm V_{rms}}$ at temperatures less than 1~K. Fig. 2 shows a sample dependence of $dI/dV$ spectra in the range of $-7~{\rm mV} < V_s < +7~{\rm mV}$, taken at $T=560$~mK in zero field. In all samples, two peaks with a width of a few mV are clearly seen both above and below $E_F$, at around $V_s = -3$~mV and $+4$~mV. Between these two peaks, shoulder-like sub-gap structures are observed at around $-1$~mV and $+2$~mV, and the local minimum of DOS lies at $E_F$. This is exactly the type of structure in the DOS most favorable to metamagnetism within a Stoner picture as described above. The overall structure of the measured tunneling spectra was independent of position, indicating that the sharp peaks are not an extrinsic impurity effect. However, the detailed distribution of spectral weight within the (intrinsic) sub-gap structure showed a small but measurable variation from site to site. Further studies will be needed to clarify this issue.

\begin{figure}
\includegraphics[width=6.5cm]{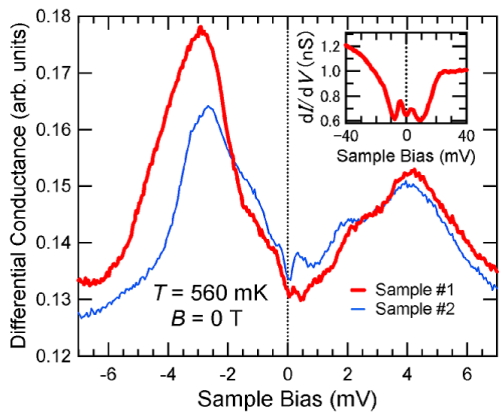}
\caption{(color online)
Sample dependence of $dI/dV$ in the range $-7~{\rm mV} < V_s < +7~{\rm mV}$ at $T = 560$~mK, $B = 0$~T (Set-point: $V_s = +7$~mV, $I_t \sim 50$~pA, Bias modulation: $V_{\rm mod} = 80 ~\mu {\rm V_{rms}}$, $f_{\rm mod} = 717$~Hz).
Sample\#1, \#2 were taken from the same batch. $dI/dV$ spectrum of sample\#1 in the range of $-40~{\rm mV} < V_s < +40~{\rm mV}$ is shown in the inset.
} 
\label{fig_2}
\end{figure}

It is interesting to compare these spectra with {\it ab initio} LSDA calculations for orthorhombic Sr$_3$Ru$_2$O$_7$~\cite{Singh}. These predict a number of sharp features in the DOS on the scale of a few hundreds meV either side of $E_F$, originating in weakly dispersing bands at the zone edge –- essentially a set of van Hove singularities. As calculated, the Fermi energy lies very close to one of these peaks. As a result, the DOS($E_F$) has the strongly enhanced value 5.0 states/ eV Ru, and the Fermi surface topology is very sensitive to small (meV) changes in the chemical potential. This is a set of conditions highly favorable for itinerant electron metamagnetism, and a theory of the metamagnetic transition in Sr$_3$Ru$_2$O$_7$ based on the rigid shift of a van Hove singularity in the DOS near $E_F$ has recently been proposed~\cite{Binz}.  

In order to establish the relationship between the peaks observed in the LDOS and metamagnetism, we investigated the detailed effects of magnetic field on the tunneling spectra. Spectra at high energies (i.e. outside the pseudo-gap) proved insensitive to field, and we therefore once again focus on the low energy regime. Repeated measurements were made in field at a {\it single} atomic site. The evolution of LDOS under magnetic field from 0 to 11 T is shown in Fig. 3(a). For fields less than the metamagnetic critical field of 8 T, the general trend is for spectral weight to be transferred asymmetrically from the peaks at $-3$~mV and $+4$~mV (which remain at roughly constant energy) into the smaller features at $-1$~mV and $+2$~mV, respectively. Above 8~T, the pattern of spectral weight transfer changes, with the local minimum in the DOS($E_F$) rapidly {\lq}{\lq}filled in{\rq}{\rq}. In contrast, topographic images taken at $V_s = +7$~mV did not show any field dependence up to 11~T. We note that, while a previous tunneling study using Sr$_3$Ru$_2$O$_7$-I-Au junction did show an increase of tunneling density of states as a function of field, it did not reveal any sharp structure in the DOS~\cite{Hooper}. We suspect that this discrepancy arises either from the quality of junction, or from matrix element effects sensitive to different sheets of the Fermi surface.

\begin{figure}
\includegraphics[width=8cm]{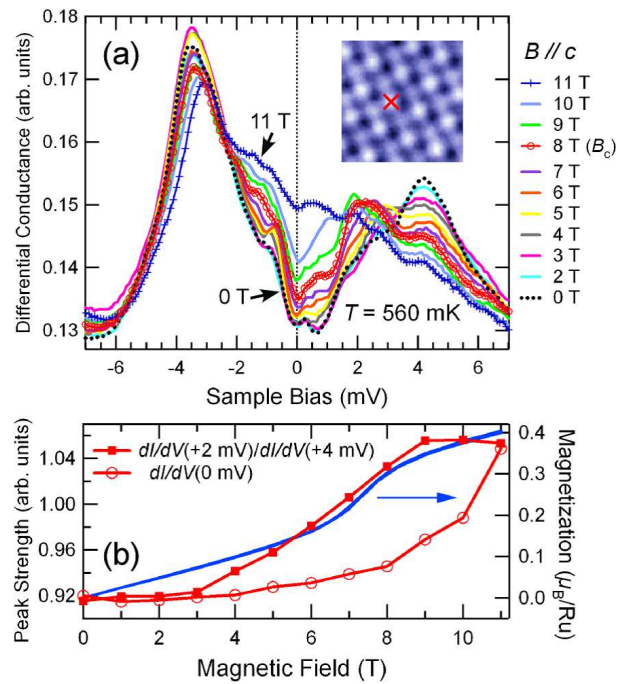}
\caption{(color online)
(a) Magnetic field dependence of $dI/dV$ spectra at $T = 560$~mK from 0 to 11 T (Set-point: $V_s = +7$~mV, $I_t \sim 50$~pA, Bias modulation: $V_{\rm mod}=200~\mu {\rm V_{rms}}$, $f_{\rm mod} = 717$~Hz).
All spectra were taken at the same position as indicated in the inset. 
(b) A ratio of $dI/dV(V_s=+2~{\rm mV})$ to $dI/dV(V_s = +4~{\rm mV})$ and normalized $dI/dV(V_s=0~{\rm mV})$ at each magnetic field calculated from (a). The magnetization curve measured at $T=2$~K is also plotted.   
}
\label{fig_3}
\end{figure}

The marked change in the nature of their evolution of spectra with field at the critical field of 8 T shows that we have captured the spectroscopic signature of a metamagnetic transition. However, the transfer of spectral weight over several meV {\it cannot} be understood simply in terms of a rigid Zeeman shift of a free electron band. In this picture, each zero-field peak splits symmetrically into separate {\lq}{\lq}up{\rq}{\rq} and {\lq}{\lq}down{\rq}{\rq} spin peaks in magnetic field. The scale of this splitting is small -- about 1~meV at 8~T -- but large enough to be resolved by our instrument. However, no such Zeeman splitting is observed, and the spectral weight is transferred over a range of energies greater than the associated change in Zeeman energy. 

Empirically, the changes in the evolution of spectra which take place at the metamagnetic transition are well characterized by taking the ratio of the height of the feature at $+2$~mV to the larger peak at $+4$~mV (Fig. 3(b)). This (arbitrary) measure of the transfer of spectral weight accurately tracks the change in magnetization at the critical field. At the same time, the weight at zero bias is a smooth, monotonically increasing function of applied field, and increases rapidly above the critical field. 

Taken together, this set of observations seems to rule out any simple {\lq}{\lq}rigid band{\rq}{\rq} theory of the metamagnetic transition in Sr$_3$Ru$_2$O$_7$. Neither our data, nor earlier dHvA measurements~\cite{Borzi} can easily be reconciled with the violent reconstruction of the Fermi surface which would accompany the chemical potential moving through a van Hove singularity. However, in a strongly correlated systems like Sr$_3$Ru$_2$O$_7$, any van Hove singularity at or near $E_F$ is likely to be split by many-body effects into a smaller set of peaks. It is therefore reasonable to suppose that the sharp low energy features in the $dI/dV$ can be understood in this way. Their evolution in magnetic field must then be governed by the same set of many-body effects, and not by a simple Zeeman splitting. 

While the spectroscopic signatures of the metamagnetic transition are very clear from these data, the $dI/dV$ spectra do not show any sharp anomaly at the critical field, unlike other (bulk) probes~\cite{Perry1,Grigera1,Kitagawa,Borzi}. A more dense data set might be required to resolve this. The absence of an anomaly at the critical field might also be due to the difficulty of probing long length-scale critical fluctuations with a purely local probe. In particular, soft paramagnon excitations (i.e. collective spin modes) associated with the metamagnetic transition would contribute to such an anomaly, but are inaccessible to a single-particle probe like tunneling spectroscopy –- except where they couple through an inelastic tunneling process~\cite{Hirjibehedin}. Soft orbital and/or lattice modes are equally hard to access. 

The surface of Sr$_3$Ru$_2$O$_7$ has been shown above to undergo an electronic ordering on an energy scale of several tens of meV, so the surface DOS might contain features absent in the bulk. However, the strong correlation between magnetization and  $dI/dV$ spectra, demonstrated in Fig. 3(b), suggests a strong admixture of the bulk in the surface DOS. It is also interesting to compare the LDOS($E_F$) under magnetic field with the enhancement of the NMR spin-lattice relaxation rate~\cite{Kitagawa} and the quasi-particle effective mass observed in dHvA experiments~\cite{Borzi}.

To date, most attempts to understand the metamagnetic transition in Sr$_3$Ru$_2$O$_7$ have been based on effective theories for the order parameter~\cite{Millis}. No microscopic model capable of describing its metamagnetic transition in the great detail revealed by STM yet exists, and our data therefore presents a new challenge to theorists. Given the many competing low energy scales it is not unlikely that charge, spin, orbital and lattice effects all play a role. However, at this stage it is too early to say whether the new electronic ordering described above is a key player in, or merely a spectator at, this transition. 

In conclusion, we have demonstrated the local signatures of criticality in the bi-layer ruthenate Sr$_3$Ru$_2$O$_7$. Local tunneling spectroscopy reveals the presence of singular features in the quasi-particle DOS on the meV-scale near $E_F$. Since the evolution of these features with magnetic field undergoes a qualitative change at the metamagnetic transition, we believe that spectroscopic evidence for metamagnetic criticality has been captured for the first time. In addition, we observed an electronic ordering at the surface in the absence of an atomic surface reconstruction. This represents another, new electronic instability inherent to strongly correlated ruthenates.

We acknowledge useful discussions with R. Matthieu, Ch. Renner. 
This work is partly supported by Grant-in-Aid for Scientific
Research from the Ministry of Education, Culture Sports, Science and
Technology of Japan.


\end{document}